\documentclass{PoS}

\title{B Semileptonic Decays at High Recoil Momentum }

\ShortTitle{B Semileptonic Decays at High Recoil Momentum}

\author{C.~T.~H.~Davies, E.~Follana and \speaker{K.~Y.~Wong}\\
        Department of Physics and Astronomy,
        University of Glasgow, Glasgow, G12 8QQ, UK}

\author{G.~P.~Lepage\\
        Laboratory of Elementary Particle Physics,
        Cornell University, Ithaca, New York 14853, USA}

\author{J.~Shigemitsu\\
        Department of Physics,
        The Ohio State University, Columbus, Ohio 43210, USA}

\abstract{We explore the possibility of studying $B\to\pi l\nu$ semileptonic
decays at large recoil momentum. Our methods include the use of a
random-wall source for the pion to reduce statistical errors, and different
smearing functions are used for the B meson to improve the overlap with the
ground state. We observe, in general, a factor of 3-4 improvement in the
signal-to-noise ratio in correlation functions if random-wall propagators
are used.}

\FullConference{The XXV International Symposium on Lattice Field Theory\\
                 July 30 - August 4 2007\\
                 Regensburg, Germany}

\begin{document}

\section{Introduction}

Precise determination of the form factors $f_{0}(q^{2})$, $f_{+}(q^{2})$
in $B\to \pi l\nu$ semileptonic decays is crucial to the determination
of the CKM-matrix element $|V_{ub}|$. Fig.~1 shows the kinematics of the
process. Lattice QCD provides a first principles nonperturbative approach
to calculate the from factors in semileptonic decays. Standard simulation
methods, however, are problematic in the low $q^{2}$ region when the pion
has large recoil momentum. Simulation results are limited to
$q^{2}\gtrsim 15\mathrm{GeV}^{2}$~\cite{Okamoto05,Dalgic06} while
experimental data spans the entire $q^{2}$ range.
Large recoil momenta are difficult for lattice calculations because
statistical errors, which are set by $E(p)-E(p=0)$, become worse when the
hadrons have large momenta. In addition, discretization errors, which are
set by $a^{2}p^{2}$, increase as the pion momentum increases.

In order to utilize all the experimental data and thereby reduce the
experimental error on $|V_{ub}|$, it is important to develop
new simulation techniques to cover the low $q^{2}$ region. One method is
to reduce the pion momentum by using a lattice frame in which the B
meson is moving in the opposite direction to the pion; to describe a
b quark with large velocity on the lattice the ``moving-NRQCD'' formalism
is used~\cite{Hashimoto96,Sloan98,Foley02,Dougall06}. In this work we
concentrate on reducing the statistical errors with
the use of a random-wall source for the light quark propagator.
In particular we explore the possibility of simulating at
$ap_{\pi}=\frac{2\pi}{L}(3,0,0)$, which corresponds to
$q^{2}\sim \textrm{10GeV}^{2}$
in the B meson rest frame. We test this on MILC coarse lattices where
discretization errors are about 8\% at this momentum; errors will be much
smaller on the fine (3\%) and super-fine (1.5\%) lattices. We also use
the Highly Improved Staggered Quark (HISQ) action~\cite{Follana03,Follana07}
for the valence light quarks.
\begin{figure}
\centering
\includegraphics[width=0.4\textwidth]{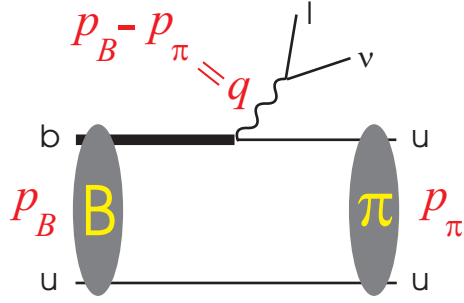}
\caption{Kinematics of $B\to\pi l\nu$ semileptonic decay.
$p_{B}$ is the momentum of the $B$ meson and $p_{\pi}$ is the
momentum of the pion, $q$ is the momentum transfer.}
\end{figure}

\section{Pion 2-Point Function}

A zero-momentum random-wall source is generated by setting the color
vector on each site of a time slice\footnote{In practice we also put the
source on a random time slice.} to a three-component random complex
unit vector $\vec{\eta}(x)$~\cite{Aubin04}. This random source is used
in matrix inversion to obtain the staggered-quark propagator $g(y,x)$
\begin{equation}
\tilde{g}(y) \equiv \sum_{x} g(y,x) \vec{\eta}(x)
= \sum_{x} M^{-1}_{y,x}\vec{\eta}(x),
\label{eq_RW}
\end{equation}
where $M_{y,x}$ is the kernel of the staggered-quark action.
It is more convenient to work in the na\"ive-quark basis. The 4-component
na\"ive-quark propagator $S(y,x)$ is given by~\cite{Follana07}
\begin{equation}
S(y,x) \equiv \big< \psi^{l}(y)\overline{\psi}^{l}(x) \big>
= g(y,x)\Omega(y)\Omega^{\dag}(x),
\end{equation}
where
\begin{equation}
\Omega(x)= \prod_{\mu=0}^{3} \left(\gamma_{\mu}\right)^{x_{\mu}}.
\end{equation}
The pion correlator in the na\"ive-quark basis becomes
{\setlength\arraycolsep{2pt}
\begin{eqnarray}
\big< J_{5}(y)J_{5}(x) \big>
& = &
\big< \big(\overline{\psi}^{l}(y)\gamma_{5}\psi^{l}(y)\big)
      \big(\overline{\psi}^{l}(x)\gamma_{5}\psi^{l}(x)\big) \big>
\nonumber \\
& = &
\mathrm{Tr} \big[
\gamma_{5}\psi^{l}(y)\overline{\psi}^{l}(x)\gamma_{5}
\psi^{l}(x)\overline{\psi}^{l}(y)
\big]
\nonumber \\
& = &
\mathrm{Tr} \big[
\gamma_{5}S(y,x)\gamma_{5}\gamma_{5}S^{\dag}(y,x)\gamma_{5} \big]
\nonumber \\
& = &
\mathrm{Tr} \big[
\Omega(y)\Omega^{\dag}(x)\Omega(x)\Omega^{\dag}(y)\big]
\mathrm{tr} \big[ \left| g(y,x) \right|^{2} \big]
\nonumber \\
& = &
4\mathrm{tr} \big[ \left| g(y,x) \right|^{2} \big]
\end{eqnarray}}
in terms of the staggered-quark propagator. Here $\mathrm{Tr}[\dots]$ is
a trace over spinor indices while $\mathrm{tr}[\dots]$ is taken over
the color indices. Therefore to construct the pion correlation function
at zero momentum we take the magnitude square of $\tilde{g}(y)$ at the
sink and sum over spatial sites, and divide by the number of sites $N$
\begin{equation}
\frac{1}{N}\sum_{y} \tilde{g}^{*}(y)\tilde{g}(y)
=
\frac{1}{N}\sum_{y,x,x'}
g^{*}(y,x')g(y,x) \vec{\eta}^{*}(x') \vec{\eta}(x).
\end{equation}
Since the averaged correlator has contributions only from where
the quark and antiquark start at the same spatial site, i.e., $x=x'$,
the random-wall source simulates many-point source, and thereby increases
the statistics.

For correlator at finite momentum $k$, an additional phase is added to
the source
\begin{equation}
\tilde{g}^{\pm}(y) \equiv \sum_{x}g(y,x)e^{\pm i\frac{k}{2}x}\vec{\eta}(x)
= \sum_{x}M^{-1}_{y,x}e^{\pm i\frac{k}{2}x}\vec{\eta}(x).
\end{equation}
In this case we multiple $\tilde{g}^{+}(y)$ and $(\tilde{g}^{-}(y))^{*}$
with an explicit insertion of $e^{ikx}$ at the sink and sum over spatial
sites. 

Fig.~2 compares the pion 2-point functions obtained with a local and a
random source at $ap_{\pi}=(0,0,0)$ and $ap_{\pi}=\frac{2\pi}{L}(3,0,0)$
on MILC coarse lattices. The advantage of using a random-wall source is
clearly demonstrated, with statistical errors about 5 times smaller
compared to the local case. To fit the 2-point functions we employ a Bayesian
technique~\cite{Lepage02} and do a multiple-exponential fit
\begin{equation}
C^{(2)}(t) = \sum_{k=0}^{N_{\pi}-1}
(-1)^{kt}a_{k}e^{-E_{\pi}^{k}t}.
\end{equation}
We use 5 exponentials $N_{\pi}=5$ (i.e., 3 normal states and 2 oscillating
states). We observe a factor of 2-3 improvement in the ground state energy
and the amplitude for the random-wall source.
\begin{figure}
\centering
\includegraphics[width=\textwidth]{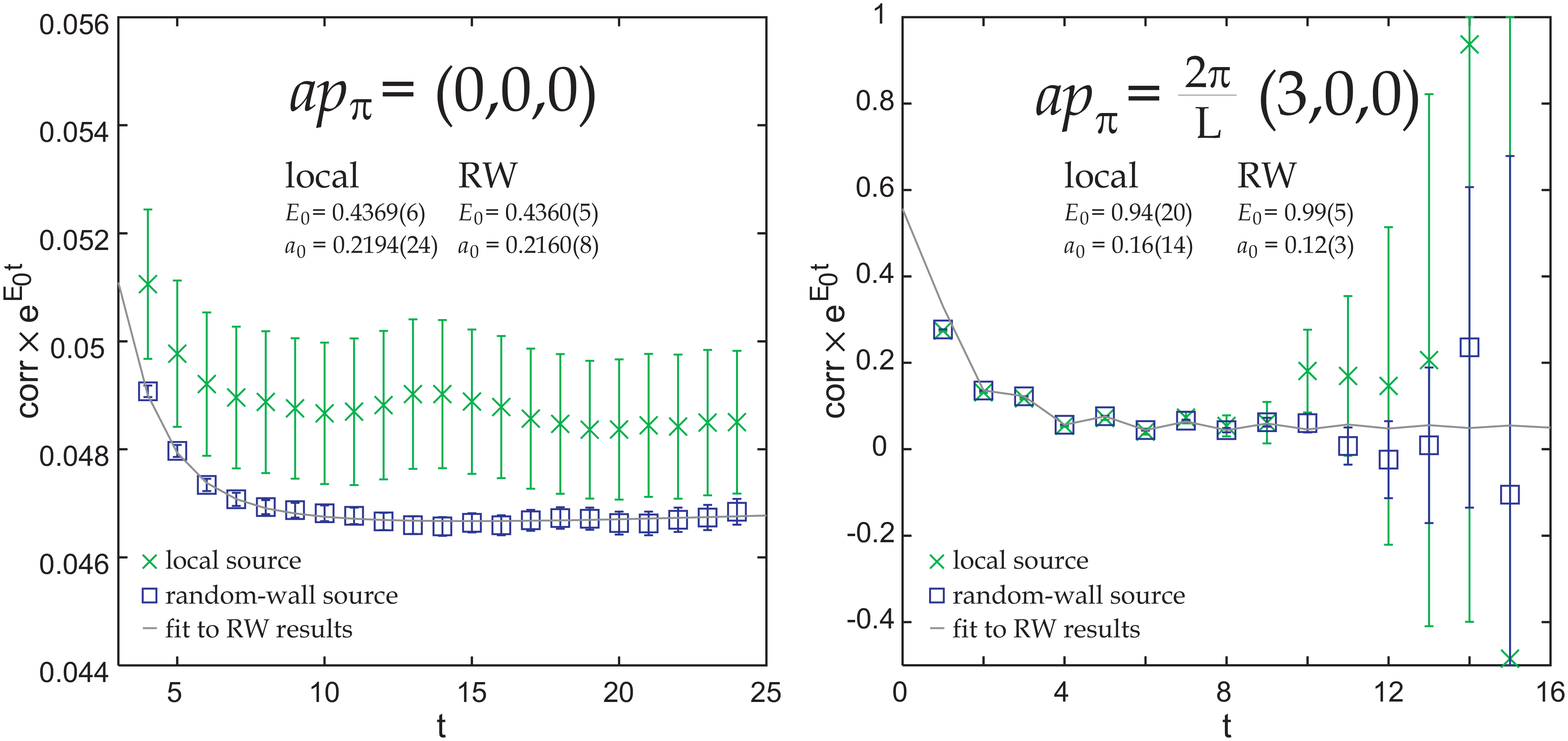}
\caption{Pion 2-point functions obtained with a local source and
a random source on MILC coarse lattices [$am_{sea}=.01/.05$,
$am_{val}=.05465$ ($\sim m_{s}$) for $ap_{\pi}=(0,0,0)$; $am_{sea}=.02/.05$,
$am_{val}=.02675$ ($\sim m_{s}/2$) for $ap_{\pi}=\frac{2\pi}{L}(3,0,0)$].
Fit results for ground state energy $E_{0}$ and amplitude $a_{0}$ are also
shown.}
\end{figure}

\section{Heavy-Light 2-Point Function}

The heavy-light correlator is
{\setlength\arraycolsep{2pt}
\begin{eqnarray}
\big< J_{5}(y)J_{5}(x) \big>
& = &
\big< \big(\overline{\psi}^{l}(y)\gamma_{5}\psi^{Q}(y)\big)
      \big(\overline{\psi}^{Q}(x)\gamma_{5}\psi^{l}(x)\big) \big>
\nonumber \\
& = &
\mathrm{Tr} \big[
\gamma_{5}\psi^{l}(x)\overline{\psi}^{l}(y)\gamma_{5}
\psi^{Q}(y)\overline{\psi}^{Q}(x)
\big]
\nonumber \\
& = &
\mathrm{Tr} \big[
\gamma_{5}\gamma_{5}S^{\dag}(y,x)\gamma_{5}\gamma_{5} G(y,x)
\big]
\nonumber \\
& = &
\mathrm{Tr} \big[ \left(\Omega^{\dag}(y) g^{*}(y,x)\right) G(y,x)\Omega(x)
\big].
\end{eqnarray}}
Here $G(y,x)$ is the heavy-quark propagator. We use NRQCD for the b quark
\begin{equation}
G(x,t+1)=
\left( 1-\frac{\delta H}{2} \right)
\left( 1-\frac{H_{0}}{2n} \right)^{n}
U_{t}^{\dag}(x)
\left( 1-\frac{H_{0}}{2n} \right)^{n}
\left( 1-\frac{\delta H}{2} \right)
G(x,t),
\end{equation}
and $H=H_{0}+\delta_{H}$ is the $O(1/M)$ improved lattice NRQCD
Hamiltonian~\cite{Lepage92,Gray05}. In order to combine with the
light-quark propagator, the heavy-quark propagator must
be initialized with the same random noise $\vec{\eta}(x)$
\begin{equation}
G(x,t=0)= \sum_{x'} \Omega(x') \Phi(|x'-x|) \vec{\eta}(x'),
\end{equation}
where $\Phi(|x'-x|)$ is the smearing function centered at $x'$.
Since the smearing function depends on the magnitude of $x'-x$ only,
$G(x,t=0)$ can be computed efficiently by applying translation to
$\Phi(|x|)$, the smearing function with center located at the origin,
when doing the summation. At the sink we multiply the heavy-quark
propagator with $\Omega^{\dag}(y) \tilde{g}^{*}(y) \times \Phi(|y|)$ and
sum over spatial sites and divide by $N$. Again only the terms with
matching random numbers survive after the ensemble average, and the
correlation function is the average of all contributions from where the
heavy quark and light quark start at the same site.

Fig.~3 shows the results obtained with a local source and a random source,
with and without smearing. The random-wall results are again more accurate,
although the improvement is not as significant as for the light-light
correlators. Gaussian smearing is used and results are shown for
local-source, local-sink [LL] (i.e., no smearing) and smeared-source,
smeared-sink [SS]. The plots show that the advantage of using
random-wall propagators decreases if smearing is used. It is therefore
crucial to fit the local and smeared results together. We fit LL, LS, SL,
SS simultaneously in a matrix fit using 5 exponentials (3 normal states
and 2 oscillating states) and
find that a smearing function of radius $ar=4$ gives the least relative
errors for the ground state energy and amplitude.
\begin{figure}
\centering
\includegraphics[width=\textwidth]{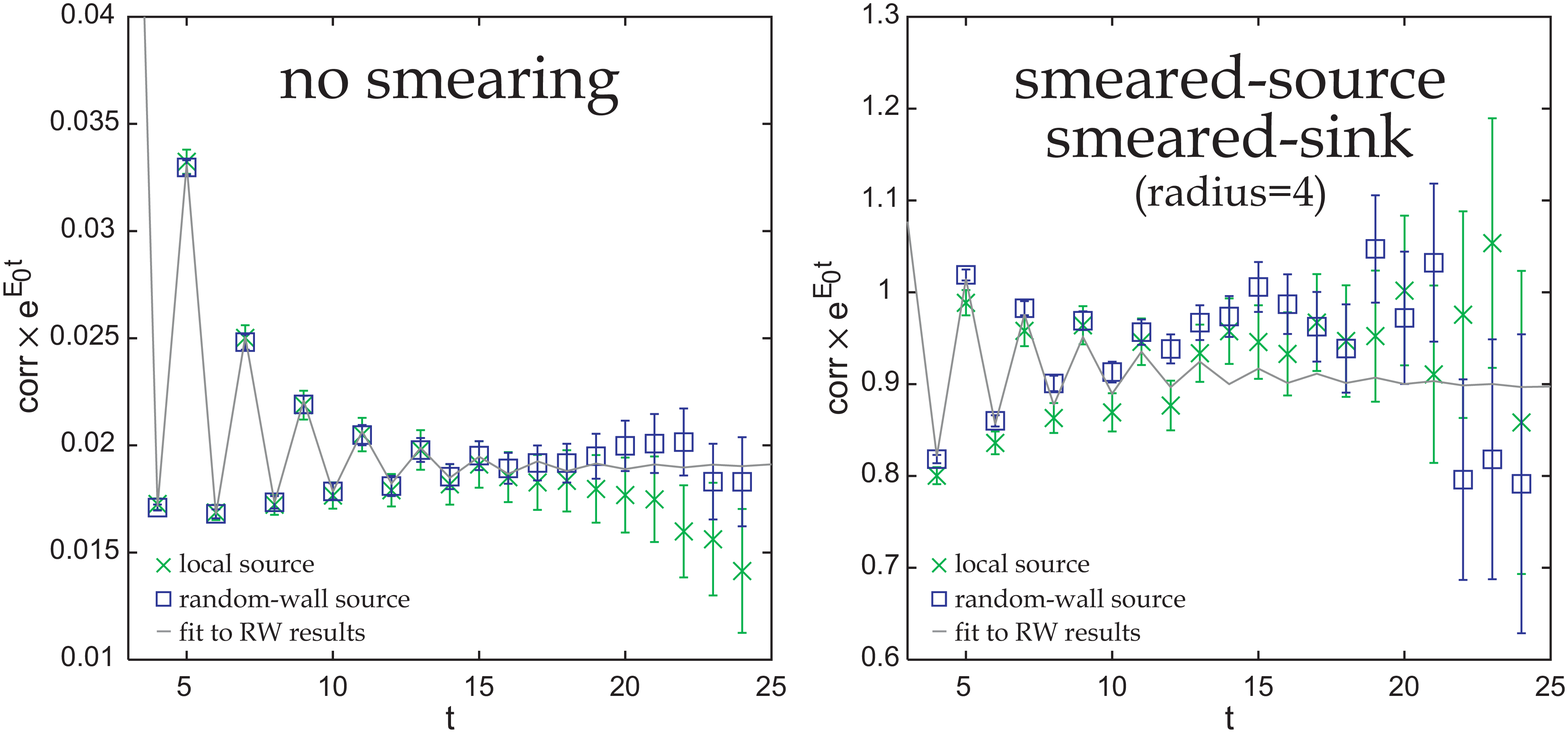}
\caption{Heavy-light 2pt-functions on MILC coarse lattices
[$am_{sea}=.01/.05$, $am_{val}=.05465$ ($\sim m_{s}$), $am_{b}=2.8$].
Gaussian smearing with radius $ar=4$ is used and results are shown for
local-source, local-sink [LL] and smeared-source, smeared-sink [SS].
A matrix fit of LL, LS, SL, SS gives [local]
$E_{0}=0.5532(15)$, $a_{0}=0.1402(10)$,
and [RW] $E_{0}=0.5559(7)$, $a_{0}=0.1424(5)$.}
\end{figure}

\section{3-Point Function}

We are interested in the 3-point function
{\setlength\arraycolsep{2pt}
\begin{eqnarray}
\big< J_{5}(y) V_{\mu}(z) J_{5}(x) \big>
& = &
\big< \big(\overline{\psi}^{l}(y)\gamma_{5}\psi^{l}(y)\big)
      \big(\overline{\psi}^{l}(z)\gamma_{\mu}\psi^{Q}(z)\big)
      \big(\overline{\psi}^{Q}(x)\gamma_{5}\psi^{l}(x)\big)
\big>
\nonumber \\
& = &
\mathrm{Tr} \big[
\gamma_{5} \psi^{l}(y) \overline{\psi}^{l}(z) \gamma_{\mu}
\psi^{Q}(z) \overline{\psi}^{Q}(x) \gamma_{5} 
\psi^{l}(x) \overline{\psi}^{l}(y) \big]
\nonumber \\
& = &
\mathrm{Tr} \big[
\gamma_{5} \gamma_{5} S^{\dag}(z,y) \gamma_{5} \gamma_{\mu}
G(z,x) \gamma_{5} S(x,y) \big]
\nonumber \\
& = &
\mathrm{Tr} \big[
\left(\Omega^{\dag}(z)g^{*}(z,y)\right) \gamma_{5}\gamma_{\mu}
G(z,x) \gamma_{5} \left(\Omega(x)g(x,y)\right) \big].
\label{eq_3pt}
\end{eqnarray}}
with $x_{0}<z_{0}<y_{0}$. To compute the current we initialize
$G(z,x)$ with $\gamma_{5}\times \Omega(x)\tilde{g}(x)$ and then propagate
the heavy quark backward\footnote{Backward means to the opposite direction
of the light-quark propagator.} in time to the current insertion point $z$,
where it
turns into a light quark; more precisely, we take the trace of the
product of $G(z,x)$ and $\Omega^{\dag}(z) \tilde{g}^{*}(z) \times
\gamma_{5}\gamma_{\mu}$ at the insertion point $z$. This setup is not
merely convenient and efficient, but the same computer code can be used
for local and random-wall propagators\footnote{In the local case we read
in local propagators $\tilde{g}(x)=\sum_{y'}g(x,y')\delta_{yy'}=g(x,y)$ and
$\tilde{g}(z)=g(z,y)$; in the random-wall case we read in the random-source
propagators Eq.~(\ref{eq_RW}).}.
We only have to divide the random-wall results by an extra factor of $N$
since there are $N$ times more contributions like the one in
Eq.~(\ref{eq_3pt}), each from a different source point $x$, to the
averaged correlator.

In Fig.~4 we plot the temporal vector currents calculated with
local and random-wall propagators for pion momenta $ap_{\pi}=(0,0,0)$ and
$ap_{\pi}=\frac{2\pi}{L}(3,0,0)$. Results again clearly show that
statistical noises can be suppressed substantially by the use of random
sources. The 3-point function has the functional form
\begin{equation}
C^{(3)}(t)=
\sum_{k=0}^{N_{\pi}-1} \sum_{l=0}^{N_{B}-1}
(-1)^{kt} (-1)^{l(T-t)}
a_{k} a_{l} v_{kl} e^{-E_{\pi}^{k}t} e^{-E_{B}^{l}(T-t)},
\end{equation}
where $T=y_{0}-x_{0}$. To extract the matrix element $v_{00}$ we fit the
3-point function, the 2-point function (Fig.~2) and the heavy-light
2-point function (Fig.~3) simultaneously using $N_{\pi}=N_{B}=5$.
Fit results are given along with the graphs. We find that random-wall
results are about 5 times more accurate compared to local case at
$ap_{\pi}=(0,0,0)$, and about a factor of 2 better
at $ap_{\pi}=\frac{2\pi}{L}(3,0,0)$.
\begin{figure}
\centering
\includegraphics[width=\textwidth]{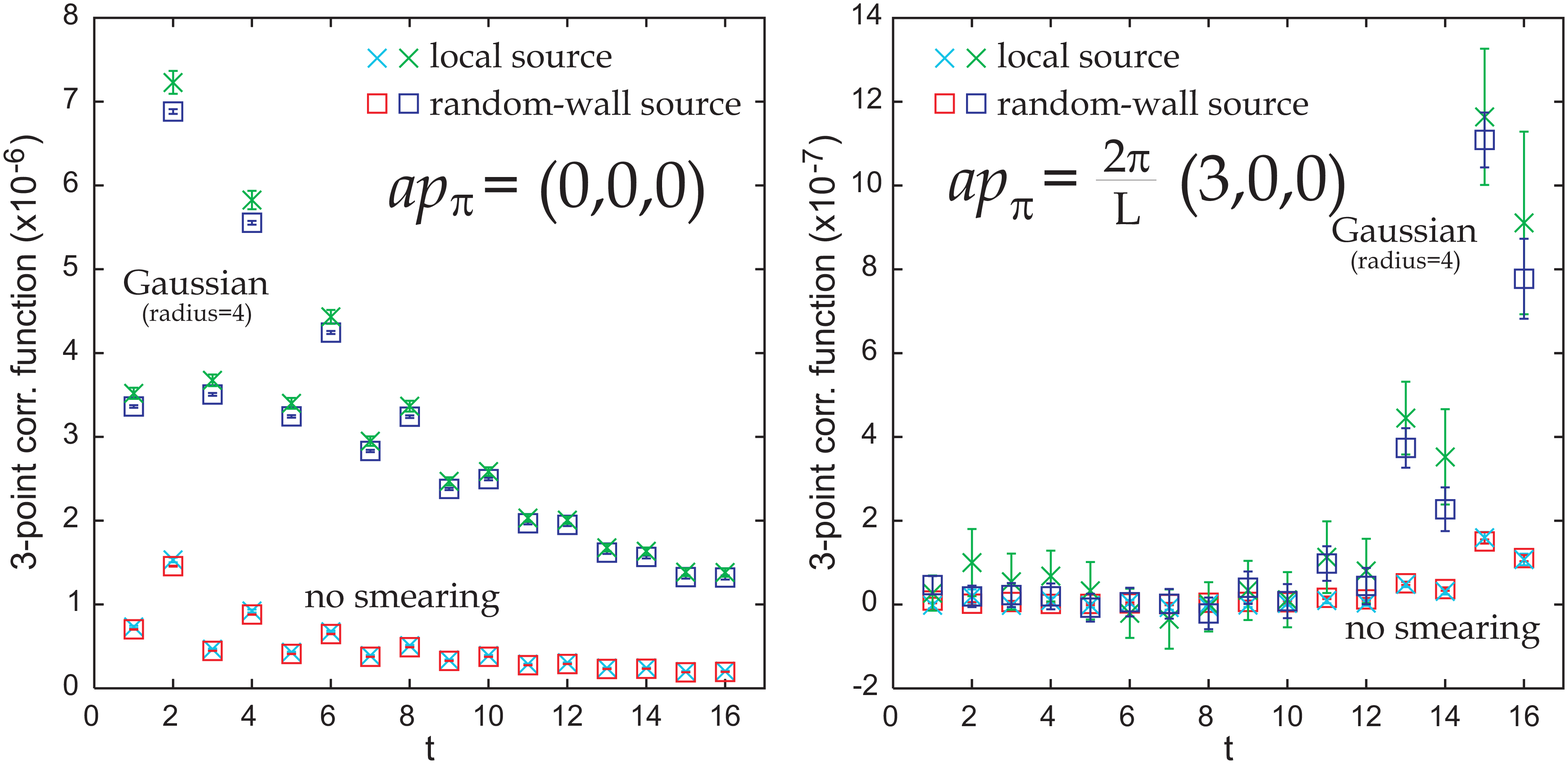}
\caption{FIG.~5. The 3-point functions (corresponding to the temporal vector
currents) obtained with local and random-wall propagators on MILC coarse
lattices [the same ensembles as in Fig.~2]. The B meson starts at $t=0$ and
the pion is at $t=16$. The random-wall results
are 3-4 more accurate compared to the results obtained
using local propagators. In particular we obtain, for no smearing,
$v_{00}=0.0605(39)$ [local], $v_{00}=0.0597(7)$ [RW] for
$ap_{\pi}=(0,0,0)$, and
$v_{00}=0.029(49)$ [local], $v_{00}=0.043(33)$ [RW] for
$ap_{\pi}=\frac{2\pi}{L}(3,0,0)$.}
\end{figure}

\vspace{0.5cm}
\section{Conclusion}

The major problem in studying semileptonic decays on the lattice is
the exponential growth of statistical errors in correlation functions
as the pion momentum increases. In this work we explored the possibility
of reducing statistical noises using random-wall light-quark propagators.
We demonstrated that correlation functions obtained with random sources
have much better signal-to-noise ratios, with statistical errors about
3-4 times smaller than those obtained with local propagators.
One should therefore improve the current lattice simulations at
small pion momenta using random-wall propagators. Encouraging
results were also obtained at pion momentum as large as
$ap_{\pi}=\frac{2\pi}{L}(3,0,0)$ (corresponding to
$q^{2}\sim 10\mathrm{GeV}^{2}$), although statistical errors
are still too large for the results to be useful. Work is in progress,
e.g. fitting correlation functions with different $T$ simultaneously,
to further improve the calculations.

\section{Acknowledgments}

This work was supported by PPARC (UK) and the DOE and NSF (USA).
We thank the MILC Collaboration for making their unquenched gauge
configurations available. The computations were done on computer
clusters at Fermilab and QCDOCX.


\begin{thebibliography}{99}

\bibitem{Okamoto05} M.~Okamoto {\it et al.},
Nucl.~Phys.~Proc.~Suppl. {\bf 140}:461-463 (2005) [hep-lat/0409116].

\bibitem{Dalgic06} E.~Dalgic {\it et al.},
Phys.~Rev.~D{\bf 73}:074502 (2006) [hep-lat/0601021].

\bibitem{Hashimoto96} S.~Hashimoto and H.~Matsufuru,
Phys.~Rev.~D{\bf 54}:4578 (1996) [hep-lat/9511027].

\bibitem{Sloan98} J.~H.~Sloan,
Nucl.~Phys.~Proc.~Suppl. {\bf 63}, 365 (1998) [hep-lat/9710061].

\bibitem{Foley02} K.~M.~Foley and G.~P.~Lepage,
Nucl.~Phys.~Proc.~Suppl. {\bf 119}, 635 (2002) [hep-lat/0209135].

\bibitem{Dougall06} A.~Dougall {\it et al.},
PoS LAT2005, 219 (2006) [hep-lat/0509108].

\bibitem{Follana03} E.~Follana {\it et al.},
Nucl.~Phys.~Proc.~Suppl.~{\bf 129}, 447 (2004) [arXiv:hep-lat/0311004].

\bibitem{Follana07} E.~Follana {\it et al.},
Phys.~Rev.~D{\bf 75}:054502 (2007) [arXiv:hep-lat/0610092].

\bibitem{Aubin04} C.~Aubin {\it et al.}, MILC Collaboration,
Phys. Rev. D{\bf 70}:114501 (2004) [hep-lat/0407028].

\bibitem{Lepage02} G.~P.~Lepage {\it et al.},
Nucl.~Phys.~Proc.~Suppl. {\bf 106}, 12 (2002) [hep-lat/0110175].

\bibitem{Lepage92} G.~P.~Lepage {\it et al.},
Phys.~Rev.~D{\bf 46}:4052 (1992) [arXiv:hep-lat/9205007].

\bibitem{Gray05} A.~Gray {\it et al.},
Phys.~Rev.~D{\bf 72}:094507 (2005) [arXiv:hep-lat/0507013].

\end{thebibliography}
\end{document}